\RequirePackage{lineno}
\documentclass[showpacs]{revtex4}    

\usepackage{graphicx}

\newcommand{\be}{\begin{equation}}
\newcommand{\ee}{\end{equation}}
\newcommand{\ba}{\begin{eqnarray}}
\newcommand{\ea}{\end{eqnarray}}
\newcommand{\ban}{\begin{eqnarray*}}
\newcommand{\ean}{\end{eqnarray*}}

\begin{document}

%\linenumbers     

\title{Measurement of event-by-event transverse momentum and multiplicity fluctuations using strongly intensive measures $\Delta[P_T, N]$ and $\Sigma[P_T, N]$ in nucleus-nucleus collisions at the CERN Super Proton Synchrotron}

\author{
% updated March 26, 2013
T.~Anticic$^{21}$, B.~Baatar$^{8}$, J.~Bartke$^{6}$, 
H.~Beck$^{9}$, L.~Betev$^{10}$, H.~Bia{\l}\-kowska$^{18}$, C.~Blume$^{9}$, 
B.~Boimska$^{18}$, J.~Book$^{9}$, M.~Botje$^{1}$,
%J.~Bracinik$^{3}$, 
P.~Bun\v{c}i\'{c}$^{10}$,
%V.~Cerny$^{3}$, 
P.~Christakoglou$^{1}$,
P.~Chung$^{17}$, O.~Chvala$^{14}$, J.~Cramer$^{15}$, V.~Eckardt$^{13}$,
%H.G.~Fischer$^{10}$,
Z.~Fodor$^{4}$, P.~Foka$^{7}$, V.~Friese$^{7}$,
M.~Ga\'zdzicki$^{9,11}$, K.~Grebieszkow$^{20}$, C.~H\"{o}hne$^{7}$,
K.~Kadija$^{21}$, A.~Karev$^{10}$, V.~Kolesnikov$^{8}$, M.~Kowalski$^{6}$, 
D.~Kresan$^{7}$,
%M.~Kreps$^{3}$, 
A.~Laszlo$^{4}$, R.~Lacey$^{17}$, M.~van~Leeuwen$^{1}$,
M.~Ma\'{c}kowiak-Paw{\l}owska$^{9,20}$, M.~Makariev$^{16}$, A.~Malakhov$^{8}$,
G.~Melkumov$^{8}$, M.~Mitrovski$^{9}$, S.~Mr\'owczy\'nski$^{11}$, 
G.~P\'{a}lla$^{4}$, A.~Panagiotou$^{2}$, 
%W.~Peryt$^{20}$, 
%M.~Pikna$^{3}$, 
J.~Pluta$^{20}$, D.~Prindle$^{15}$,
F.~P\"{u}hlhofer$^{12}$, R.~Renfordt$^{9}$, C.~Roland$^{5}$, G.~Roland$^{5}$,
M. Rybczy\'nski$^{11}$, A.~Rybicki$^{6}$, A.~Sandoval$^{7}$, 
A. Rustamov$^{9}$,
N.~Schmitz$^{13}$, T.~Schuster$^{9}$, P.~Seyboth$^{13}$, F.~Sikl\'{e}r$^{4}$, 
%B.~Sitar$^{3}$, 
E.~Skrzypczak$^{19}$, M.~S{\l}\-odkowski$^{20}$, G.~Stefanek$^{11}$, R.~Stock$^{9}$, 
H.~Str\"{o}bele$^{9}$, T.~Susa$^{21}$, M.~Szuba$^{20}$, 
D.~Varga$^{3}$, M.~Vassiliou$^{2}$,
G.~Veres$^{4}$, G.~Vesztergombi$^{4}$, D.~Vrani\'{c}$^{7}$,
%S.~Wenig$^{10}$,
Z.~W{\l}odarczyk$^{11}$, A.~Wojtaszek-Szwarc$^{11}$ 
\newline
(NA49 Collaboration)
}

\vspace{0.5cm}
\affiliation{
$^{1}$ NIKHEF, Amsterdam, Netherlands. \\
$^{2}$ Department of Physics, University of Athens, Athens, Greece.\\
%$^{3}$ Comenius University, Bratislava, Slovakia.\\
$^{3}$ E\"otv\"os Lor\'ant University, Budapest, Hungary \\
$^{4}$ Wigner Research Center for Physics, Hungarian Academy of Sciences, Budapest, Hungary.\\
$^{5}$ MIT, Cambridge, Massachusetts, USA.\\
$^{6}$ H.~Niewodnicza\'nski Institute of Nuclear Physics, Polish Academy of Sciences, Cracow, Poland.\\
$^{7}$ GSI Helmholtzzentrum f\"{u}r Schwerionenforschung GmbH, Darmstadt, Germany.\\
$^{8}$ Joint Institute for Nuclear Research, Dubna, Russia.\\
$^{9}$ Fachbereich Physik der Universit\"{a}t, Frankfurt, Germany.\\
$^{10}$ CERN, Geneva, Switzerland.\\
$^{11}$ Institute of Physics, Jan Kochanowski University, Kielce, Poland.\\
$^{12}$ Fachbereich Physik der Universit\"{a}t, Marburg, Germany.\\
$^{13}$ Max-Planck-Institut f\"{u}r Physik, Munich, Germany.\\
$^{14}$ Institute of Particle and Nuclear Physics, Charles University, Prague, Czech Republic.\\
$^{15}$ Nuclear Physics Laboratory, University of Washington, Seattle, Washington, USA.\\
$^{16}$ Institute for Nuclear Research and Nuclear Energy, BAS, Sofia, Bulgaria.\\
$^{17}$ Department of Chemistry, Stony Brook University (SUNYSB), Stony Brook, New York, USA.\\
$^{18}$ National Center for Nuclear Research, Warsaw, Poland.\\
$^{19}$ Institute for Experimental Physics, University of Warsaw, Warsaw, Poland.\\
$^{20}$ Faculty of Physics, Warsaw University of Technology, Warsaw, Poland.\\
$^{21}$ Rudjer Boskovic Institute, Zagreb, Croatia.\\
}

%%%%%%%%%%%%%%%%%%%%%%%%%%%%%%%%%%%%%%%%%%%%%%%%%%%%%%%%%%%%%%%%%%%%%%%%%%%%%%%%%%%%%%%%%%%

\date{\today}

\begin{abstract}
Results from the NA49 experiment at the CERN SPS are presented on event-by-event transverse momentum and multiplicity fluctuations of charged particles, produced at forward rapidities in central Pb+Pb interactions at beam momenta 20$A$, 30$A$, 40$A$, 80$A$, and 158$A$ GeV/c, as well as in systems of different size ($p+p$, C+C, Si+Si, and Pb+Pb) at 158$A$ GeV/c. This publication extends the previous NA49 measurements of the strongly intensive measure $\Phi_{p_T}$ by a study of the recently proposed strongly intensive measures of fluctuations $\Delta[P_T, N]$ and $\Sigma[P_T, N]$. In the explored kinematic region transverse momentum and multiplicity fluctuations show no significant energy dependence in the SPS energy range. However, a remarkable system size dependence is observed for both $\Delta[P_T, N]$ and $\Sigma[P_T, N]$, with the largest values measured in peripheral Pb+Pb interactions. The results are compared with NA61/SHINE measurements in $p+p$ collisions, as well as with predictions of the UrQMD and EPOS models.
\end{abstract}

\vspace{0.3cm}


%25.75.-q Relativistic heavy-ion collisions
%25.75.Gz Particle correlations

\maketitle

%%%%%%%%%%%%%%%%%%%%%%%%%%%%%%%%%%%%%%%%%%%%%%%%%%%%%%%%%%%%%%%%%%%%
\section{Introduction and motivation}
%%%%%%%%%%%%%%%%%%%%%%%%%%%%%%%%%%%%%%%%%%%%%%%%%%%%%%%%%%%%%%%%%%%%

Ultra-relativistic heavy ion collisions are studied mainly to understand the properties of strongly interacting matter under extreme conditions of high energy densities when the creation of the quark-gluon plasma (QGP) is expected. The results obtained in a broad collision energy range by experiments at the Super Proton Synchrotron (SPS) at CERN, the Relativistic Heavy Ion Collider (RHIC) at BNL, and at the Large Hadron Collider (LHC) at CERN indeed suggest that in collisions of heavy nuclei such a state with sub-hadronic degrees of freedom appears when the system is sufficiently hot and dense.

The phase diagram of strongly interacting matter is most often presented in terms of temperature ($T$) and baryochemical potential ($\mu_B$), which reflects net-baryon density. It is commonly believed that for large values of $\mu_B$ the phase transition is of the first order and turns into a rapid but continuous transition (cross-over) for low $\mu_B$ values. A critical point of second order (CP) separates these two regions. The phase diagram can be scanned by varying the energy and the size of the colliding nuclei and the CP is believed to cause a maximum of fluctuations in the measured final state particles. More specifically, the CP is expected to lead not only to non-Poissonian distributions of event quantities like multiplicities or average transverse momentum~\cite{Stephanov:1999zu, Athanasiou:2010kw}, but also to intermittent behavior of low-mass $\pi^+\pi^-$ pair and proton production with power-law exponents calculable in QCD~\cite{Antoniou:2005am, Antoniou:2006zb}.

The NA49 experiment at the CERN~SPS~\cite{Afanasev:1999iu} pioneered the exploration of the phase diagram by an energy scan for central Pb+Pb collisions in the range 20$A$ to 158$A$~GeV ($\sqrt{s_{NN}}$~=~6.3--17.3~GeV), as well as a system size scan at the top SPS energy of 158$A$~GeV. Evidence was found~\cite{Afanasiev:2002mx, Alt:2007aa} that quark/gluon deconfinement sets in at a beam energy of about 30$A$~GeV. Thus the SPS energy range is a region where the CP could be located. 
At present the search for the critical point is vigorously pursued by the NA61/SHINE collaboration at the SPS~\cite{shine} and by the beam energy scan program BES at RHIC~\cite{Aggarwal:2010cw}.

The NA49 experiment already measured multiplicity fluctuations in terms of the scaled variance $\omega$ of the distribution of event multiplicity $N$~\cite{Alt:2006jr, Alt:2007jq} and event-by-event fluctuations of the transverse momentum of the particles employing the strongly intensive measure $\Phi_{p_T}$~\cite{Anticic:2003fd, Anticic:2008aa}. The present paper reports a continuation of this NA49 study by analyzing two new strongly intensive measures of event-by-event transverse momentum and multiplicity fluctuations, $\Delta[P_T, N]$ and $\Sigma[P_T, N]$ \cite{Gorenstein:2011vq, Gazdzicki:2013ana}. These measures are dimensionless and have scales given by two reference values, namely they are equal to zero in case of no fluctuations and one in case of independent particle production. Unlike $\Phi_{p_T}$ they allow to classify the strength of fluctuations on a common scale.

This paper is organized as follows. In Sec.~\ref{strong_meas} the new strongly intensive measures of fluctuations $\Delta[P_T, N]$ and $\Sigma[P_T, N]$ are introduced and briefly discussed. Data sets, acceptance used for this analysis, detector effects, and systematic uncertainty estimates are discussed in Sec.~\ref{datasets}. The NA49 results on the energy and system size dependences of transverse momentum and multiplicity fluctuations quantified by the new measures are presented and discussed in Sec.~\ref{results}. A summary closes the paper.

%%%%%%%%%%%%%%%%%%%%%%%%%%%%%%%%%%%%%%%%%%%%%%%%%%%%%%%%%%%%%%%%%%%%%%%%%%%%%%%%%%%%%%%%%%%
\section{Strongly intensive measures of transverse momentum and multiplicity fluctuations}
\label{strong_meas}
%%%%%%%%%%%%%%%%%%%%%%%%%%%%%%%%%%%%%%%%%%%%%%%%%%%%%%%%%%%%%%%%%%%%%%%%%%%%%%%%%%%%%%%%%%%

In thermodynamics {\it extensive} quantities are those which are proportional to the system volume. Examples of extensive quantities in this case are the mean multiplicity or the variance of the multiplicity distribution. In contrast, {\it intensive} quantities are defined such that they do not depend on the volume of the system. 
It was shown \cite{Gorenstein:2011vq} that the ratio of two extensive quantities is an intensive quantity, and therefore, the ratio of mean multiplicities, as well as the commonly used scaled variance of the distribution of the multiplicity $N$, $\omega[N] = (\langle N^2 \rangle - \langle N \rangle ^2)/ \langle N \rangle$, are intensive measures. Finally, one can define a class of {\it strongly intensive} quantities which depend neither on the volume of the system nor on the volume fluctuations within the event ensemble. Such quantities can be truly attractive when studying heavy ion collisions, where the volume of the produced matter cannot be fixed and may change significantly from one event to another. Examples of strongly intensive quantities are mean multiplicity ratios, the $\Phi$ measure of fluctuations \cite{Gazdzicki:1992ri}, and the recently introduced $\Delta$ and $\Sigma$ measures of fluctuations \cite{Gorenstein:2011vq, Gazdzicki:2013ana}. In fact, it was shown \cite{Gorenstein:2011vq} that there are at least two families of strongly intensive measures: $\Delta$ and $\Sigma$. The previously introduced measure $\Phi$ is a member of the $\Sigma$-family.

In nucleus-nucleus collisions the volume is expected to vary from event to event and these changes are impossible to eliminate fully. Thus, the strongly intensive quantities allow, at least partly, to overcome the problem of volume fluctuations.  
Generally, the $\Delta$ and $\Sigma$ measures can be calculated for {\it any} two extensive quantities $A$ and $B$. In this paper $B$ is taken to be the {\it accepted} particle multiplicity, $N$ ($B \equiv N$), and $A$ the sum of their transverse momenta $P_T$ ($A \equiv P_T = \sum_{i=1}^{N} p_{T_{i}}$, the summation runs over the transverse momenta $p_{T_{i}}$ of all {\it accepted} particles in a given event). Following Refs.~\cite{Gorenstein:2011vq, Gazdzicki:2013ana} the quantities $\Delta[P_T, N]$ and $\Sigma[P_T, N]$ are defined as:

\begin{equation}
\label{delta_eq}
\Delta[P_T, N]= \frac {1}{\langle N \rangle \omega(p_{T})} [ \langle N \rangle \omega[P_T] - \langle P_T \rangle \omega[N] ] 
\end{equation}

\noindent
and

\begin{equation}
\label{sigma_eq}
\Sigma[P_T, N] = \frac {1}{\langle N \rangle \omega(p_{T})} [\langle N \rangle \omega[P_T] 
+ \langle P_T \rangle \omega[N] - 2 ( \langle P_T N \rangle - \langle P_T \rangle \langle N \rangle ) ],
\end{equation}

\noindent
where: 
\begin{equation}
\omega[P_T] = \frac {\langle {P_T}^2 \rangle - {\langle P_T \rangle}^2 }  {\langle P_T \rangle}  
\end{equation}
and  
\begin{equation}
\omega[N] = \frac {\langle N^2 \rangle - {\langle N \rangle}^2 } {\langle N \rangle}
\end{equation}

\noindent
are the scaled variances of the two fluctuating extensive quantities $P_T$ and $N$, respectively. 
The brackets $\langle ... \rangle$ represent averaging over events. The quantity $\omega(p_{T})$ is the scaled variance of the {\it inclusive} $p_{T}$ distribution (all accepted particles and events are used): 

\begin{equation}
\omega(p_{T}) = \frac {\overline{p_T^{2}} - \overline{p_T}^2 } {\overline{p_T}} 
\end{equation}

Equations~(\ref{delta_eq}) and (\ref{sigma_eq}) can be used only when assuming that $\omega(p_{T})$ is not equal to zero. There is an important difference between the $\Delta[P_T, N]$ and $\Sigma[P_T, N]$ measures. Only the first two moments: $\langle P_T \rangle$, $\langle N \rangle$, and $\langle {P_T}^2 \rangle$, $\langle N^2 \rangle$ are required to calculate  $\Delta[P_T, N]$, whereas $\Sigma[P_T, N]$ includes also the correlation term $\langle P_T N \rangle - \langle P_T \rangle \langle N \rangle$. Therefore $\Delta[P_T, N]$ and $\Sigma[P_T, N]$ can be sensitive to various physics effects in different ways. In Ref.~\cite{Gorenstein:2011vq} all strongly intensive quantities containing the correlation term are named the $\Sigma$ family, whereas those based only on mean values and variances the $\Delta$ family. As already mentioned, the previously studied \cite{Anticic:2003fd, Anticic:2008aa} measure $\Phi_{p_T}$ belongs to the $\Sigma$ family and obeys the relation:

\begin{equation}
\Phi_{p_T} = \sqrt{\overline{p_T} \omega(p_T)} [\sqrt{\Sigma[P_T, N]}-1] 
\end{equation}

With the normalization of $\Delta$ and $\Sigma$ proposed in Ref.~\cite{Gazdzicki:2013ana} these quantities are dimensionless and have a common scale making possible a quantitative comparison of fluctuations of different, in general dimensional, extensive quantities. The basic properties of the $\Delta[P_T, N]$ and $\Sigma[P_T, N]$ measures are the following:

\begin{enumerate}
\item {\it Absence of fluctuations}. In the absence of event-by-event fluctuations ($N = const.$, $P_T = const.$) the values of $\Delta[P_T, N]$ and $\Sigma[P_T, N]$ are equal to zero.

\item {\it Independent Particle Model} (IPM). If the system consists of particles that are emitted independently from each other (no inter-particle correlations) $\Delta[P_T, N]$ and $\Sigma[P_T, N]$ are equal to one. For this case $\Phi_{p_{T}}$ vanishes.

\item {\it Model of Independent Sources} (MIS). When particles are emitted by a number ($N_S$) of identical sources, which are independent of each other and $P(N_S)$ is the distribution of this number, then $\Delta[P_T, N](N_S)$ and $\Sigma[P_T, N](N_S)$ are independent of $N_S$ (intensive measures) and of its distribution $P(N_S)$ (strongly intensive measures). The $\Phi_{p_{T}}$ measure has the same property. An example of MIS is the Wounded Nucleon Model (WNM) \cite{Bialas:1976ed}, where $N_S \equiv N_W$ (number of wounded nucleons). Another example is a model where nucleus-nucleus ($A+A$) collisions are an incoherent superposition of many independent nucleon-nucleon ($N+N$) interactions. For these cases all three fluctuation measures, namely $\Phi_{p_{T}}$, $\Delta[P_T, N]$ and $\Sigma[P_T, N]$, are independent of the number of sources (and therefore insensitive to the centrality of the collisions) and have the same values for $A+A$ and $N+N$ collisions.  

\end{enumerate}

The measures $\Delta[P_T, N]$ and $\Sigma[P_T, N]$ have similar advantages to $\omega[N]$. 
Like $\omega[N]$ they have two reference values, namely $\omega[N]$ equals zero when the multiplicity is constant from event to event and one for a Poisson multiplicity distribution. Therefore one can judge whether fluctuations are large (~$ > 1$~ ) or small (~$ < 1$~ ) compared to independent particle production. However, 
$\omega[N]$ is not a strongly intensive quantity, and in the MIS one finds $\omega[N](N_s$ sources) = $\omega[N]$ (1 source) + $\langle n \rangle \omega[N_S]$, where $\langle n \rangle$ is the mean multiplicity of particles from a single source and $\omega[N_S]$ represents fluctuations of $N_S$. 

A comparison of the properties of $\Delta[P_T, N]$, $\Sigma[P_T, N]$, and $\Phi_{p_{T}}$ is presented in Table \ref{measures_tab}.  

The quantities $\Delta[P_T, N]$ and $\Sigma[P_T, N]$ were studied in several models. The results of simulations of the IPM, the MIS, source-by-source temperature fluctuations (example of MIS), event-by-event (global) temperature fluctuations, and anti-correlation between $P_T/N$ and $N$ were studied in Ref.~\cite{Gorenstein:2013nea}. Predictions from the UrQMD model on the system size and on the energy dependence of $\Delta[P_T, N]$ and $\Sigma[P_T, N]$ are shown in Ref.~\cite{Gazdzicki:2013ana}. Finally, the effects of quantum statistics were discussed in Ref.~\cite{Gorenstein:2013iua}. The general conclusion is that $\Delta[P_T, N]$ and $\Sigma[P_T, N]$ measure deviations from the superposition model in different ways. Therefore, the interpretation of the experimental results may benefit from a simultaneous measurement of both quantities.

\begin{table}[h]
\begin{center}
\begin{tabular}{|c|c|c|c|c|}
\hline
 & Unit & No fluctuations & Independent Particle Model & Model of Independent Sources \cr
\hline
\hline
$\Phi_{p_T}$  &   MeV/c       &  $\Phi_{p_T}=- \sqrt { \overline{p_T} \omega(p_T)}$               & $\Phi_{p_T}=0$ &  independent of $N_S$ and $P(N_S)$;  \cr
 & & & & $\Phi_{p_T}$($N_S$ sources) = $\Phi_{p_T}$(1 source)  \cr
\hline
$\Delta[P_T, N]$ & dimensionless & $\Delta[P_T, N]=0$ & $\Delta[P_T, N]=1$  &  independent of $N_S$ and $P(N_S)$; \cr
& & & & $\Delta[P_T, N]$($N_S$ sources) = $\Delta[P_T, N]$(1 source)  \cr
\hline
$\Sigma[P_T, N]$ & dimensionless & $\Sigma[P_T, N]=0$ & $\Sigma[P_T, N]=1$ & independent of $N_S$ and $P(N_S)$; \cr
& & & & $\Sigma[P_T, N]$($N_S$ sources) = $\Sigma[P_T, N]$(1 source)  \cr
\hline
\end{tabular}
\end{center}
\caption {Properties of $\Phi_{p_T}$, $\Delta[P_T, N]$, and $\Sigma[P_T, N]$ in the absence of fluctuations ($N=const.$, $P_T=const.$), in the Independent Particle Model and in the Model of Independent Sources.}
\label{measures_tab}
\end{table}

%%%%%%%%%%%%%%%%%%%%%%%%%%%%%%%%%%%%%%%%%%%%%%%%%%%%%%%%%%%%%%%%%%%%
\section{Data selection and analysis}
\label{datasets}
%%%%%%%%%%%%%%%%%%%%%%%%%%%%%%%%%%%%%%%%%%%%%%%%%%%%%%%%%%%%%%%%%%%%

The data used for the analysis, event and particle selection criteria, uncertainty estimates and corrections are described in the previous publications of NA49 \cite{Anticic:2003fd, Anticic:2008aa} on the measure $\Phi_{p_T}$. Here we only recall the key points. 

The analysis of the energy dependence of transverse momentum and multiplicity fluctuations uses samples of Pb+Pb collisions at 20$A$, 30$A$, 40$A$, 80$A$ and 158$A$ GeV/c beam momenta (center of mass energies from 6.3 to 17.3~GeV per $N+N$ pair) for which the 7.2\% most central reactions were selected. The analysis of the system size dependence is based on samples of $p+p$, semi-central C+C, semi-central Si+Si, and minimum bias and central Pb+Pb collisions at 158$A$ GeV/c beam momentum. Minimum bias Pb+Pb events were divided into six centrality bins (see Ref.~\cite{Anticic:2003fd} for details) but due to a trigger bias the most peripheral bin (6) is not used in the current analysis. For each bin of centrality the mean number of wounded nucleons $\langle N_{W} \rangle$ was determined by use of the Glauber model and the VENUS event generator~\cite{Werner:1993uh} (see Ref.~\cite{Anticic:2003fd}).

Tracks were restricted to the transverse momentum region $0.005 < p_T < 1.5$ GeV/c. For the study of energy dependence the forward rapidity range $1.1 < y^*_{\pi} < 2.6$ was selected, where $y^*_{\pi}$ is the particle rapidity calculated in the center-of-mass reference system. For the study of system size dependence at 158$A$ GeV/c the rapidity was calculated in the laboratory reference system and restricted to the region 4.0 $< y_{\pi} <$ 5.5~\cite{Anticic:2003fd} (it approximately corresponds to $1.1 < y^*_{\pi} < 2.6$). As track-by-track identification was not applied, the rapidities were calculated assuming the pion mass for all particles. For the energy scan an additional cut on the rapidity $y_{p}^{*}$ calculated with the proton mass was applied ($y_{p}^{*} < y^{*}_{beam}-0.5$) \cite{Anticic:2008aa, Grebieszkow:2007bv}. This excludes the projectile rapidity domain where particles may be contaminated by e.g. elastically scattered or diffractively produced protons.

\begin{figure}
\centering
\includegraphics[width=0.4\textwidth]{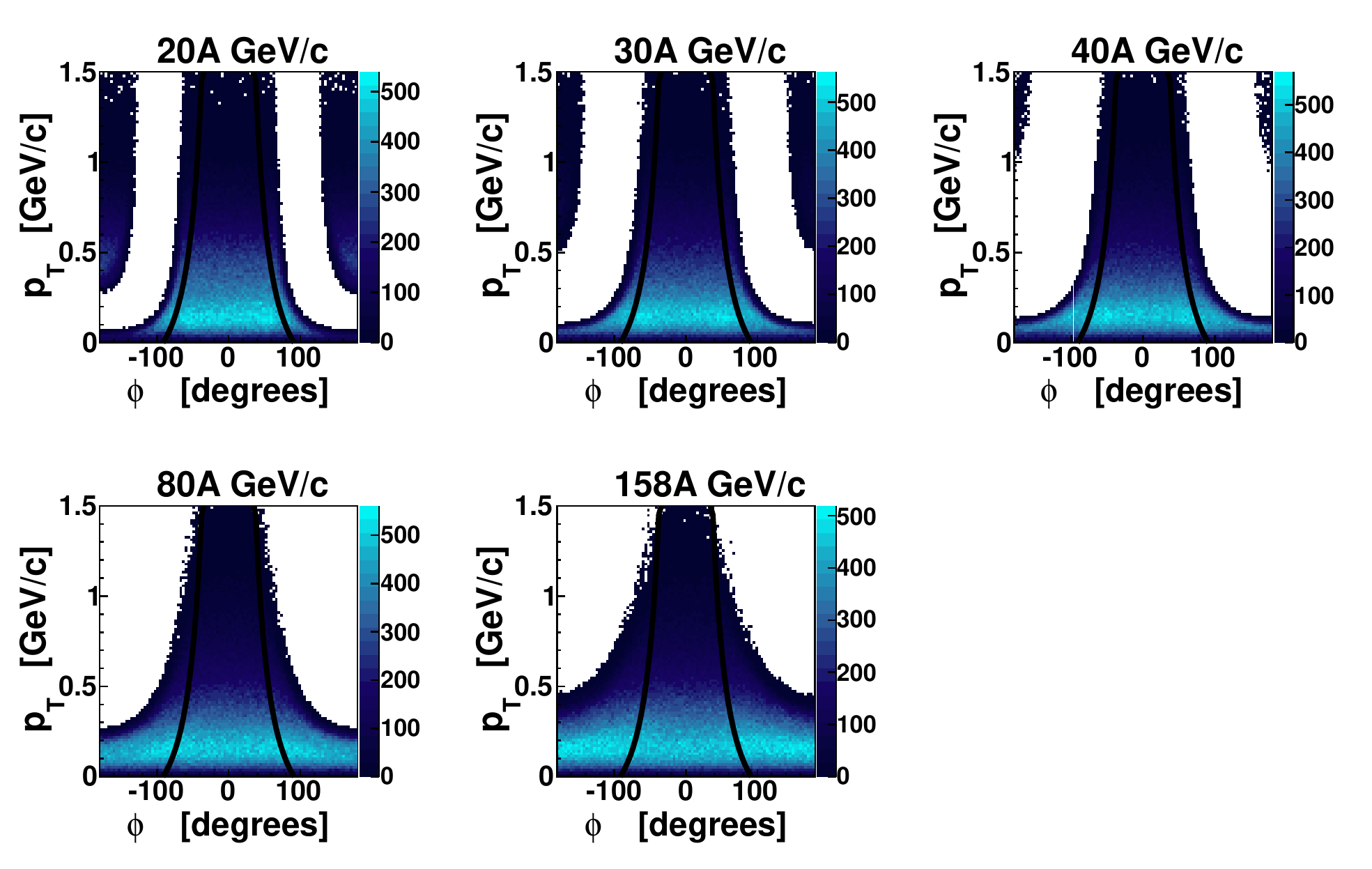}
\includegraphics[width=0.5\textwidth]{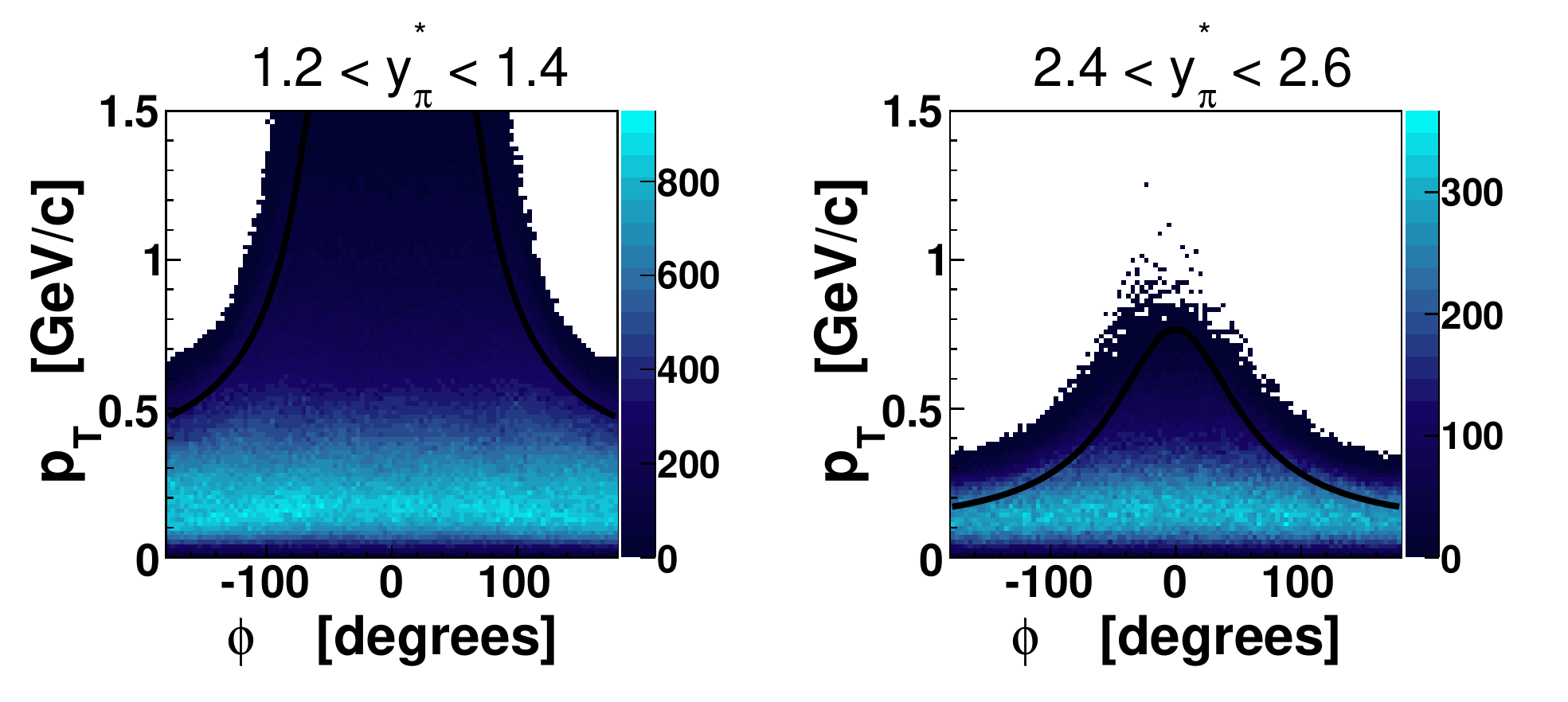}
\caption[]{(Color online) Examples of NA49 ($\phi,p_{T}$) acceptances of charged particles with the azimuthal angle of negatively charged particles reflected (see Ref.~\cite{Anticic:2008aa} for details). The solid lines represent the analytical parametrization of acceptance used for further analysis. Left: acceptance used for the energy scan of $p_T$ and $N$ fluctuations, example for $2.0 < y^{*}_{\pi} <2.2$. Right: acceptance used for the system size dependence of $p_T$ and $N$ fluctuations, examples for $1.2 < y^{*}_{\pi} <1.4$ and $2.4 < y^{*}_{\pi} <2.6$. Additional cut on $y_{p}^{*}$ (see the text) not included. Figure reproduced from Refs.~\cite{Anticic:2003fd, Anticic:2008aa} where parametrizations of the curves can be found.} 
\label{acc_sys_energy}
\end{figure}

The acceptance of azimuthal angle $\phi$ was chosen differently for the study of energy and system size dependence (Fig.~\ref{acc_sys_energy}). For the energy scan a common region of azimuthal angle was selected for all five energies (only particles within the solid curves in Fig.~\ref{acc_sys_energy}~(left) were retained), whereas a wider range was used at 158$A$~GeV/c for the system size study (see Fig.~\ref{acc_sys_energy}~(right)). Together with the track quality criteria and rapidity cuts this results in using only about 5 \% respectively 20 \% of all charged particles produced in the reactions.

\begin{figure}
\centering
\includegraphics[width=0.4\textwidth]{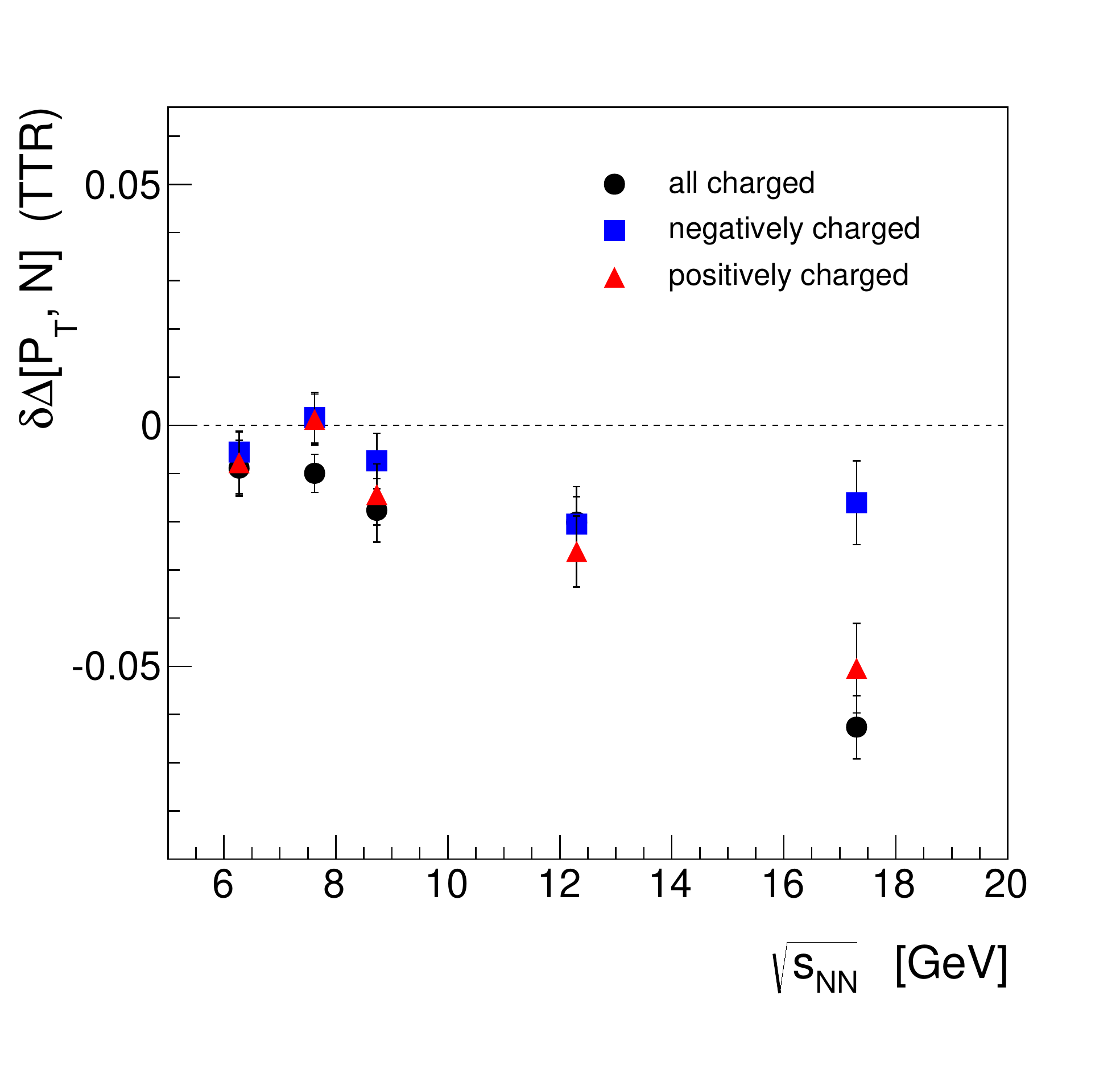}
\includegraphics[width=0.4\textwidth]{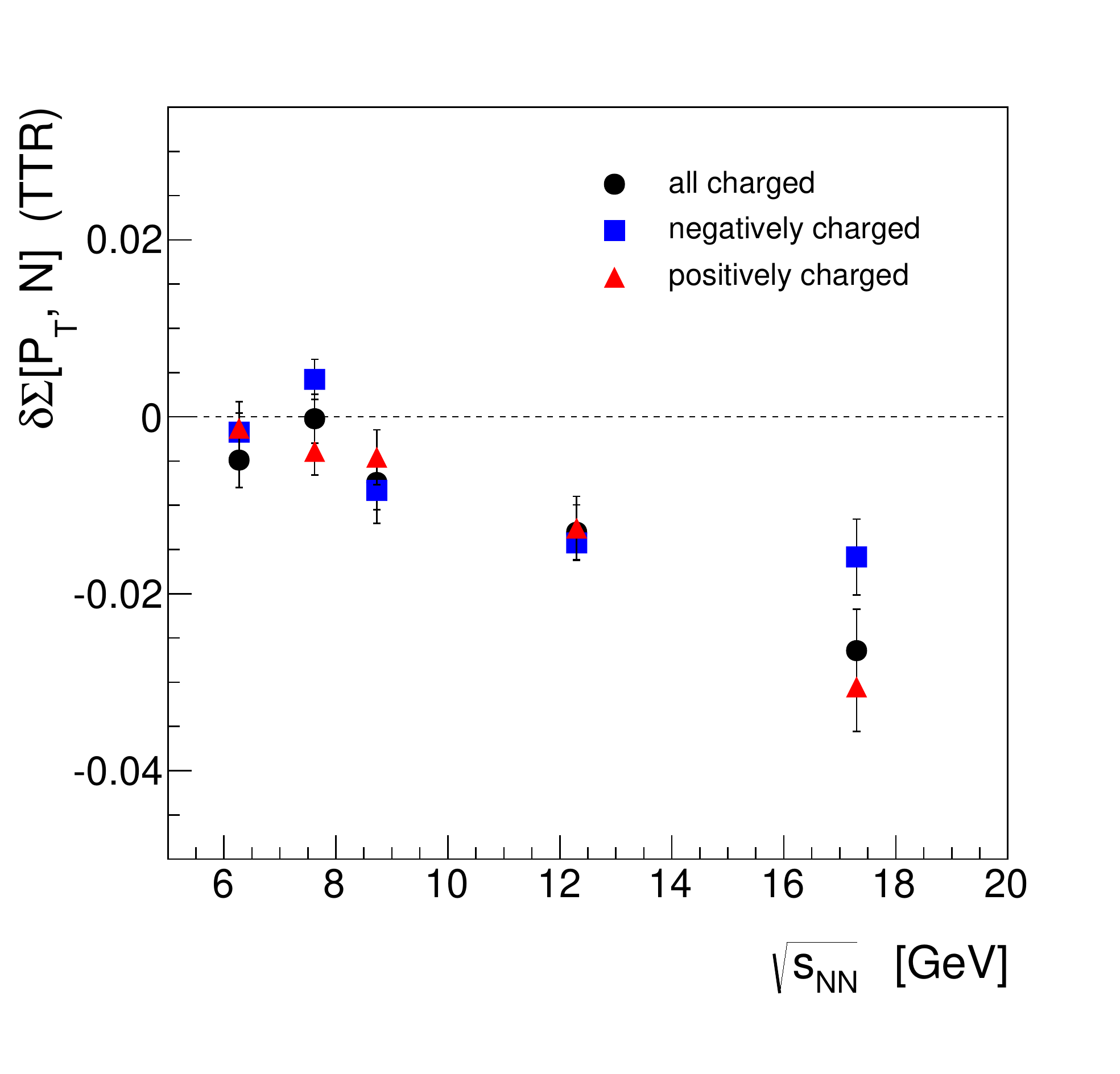}
\caption[]{(Color online) Additive corrections $\delta\Delta[P_T, N]$ (left) and $\delta\Sigma[P_T, N]$ (right) for limited two track resolution in the 7.2\% most central Pb+Pb events at 20$A$--158$A$ GeV/c. Estimates for positively charged, negatively charged and all charged particles are distinguished by different markers (see legend).}
\label{delta_sigma_energy_ttr}
\end{figure}

\begin{figure}
\centering
\includegraphics[width=0.4\textwidth]{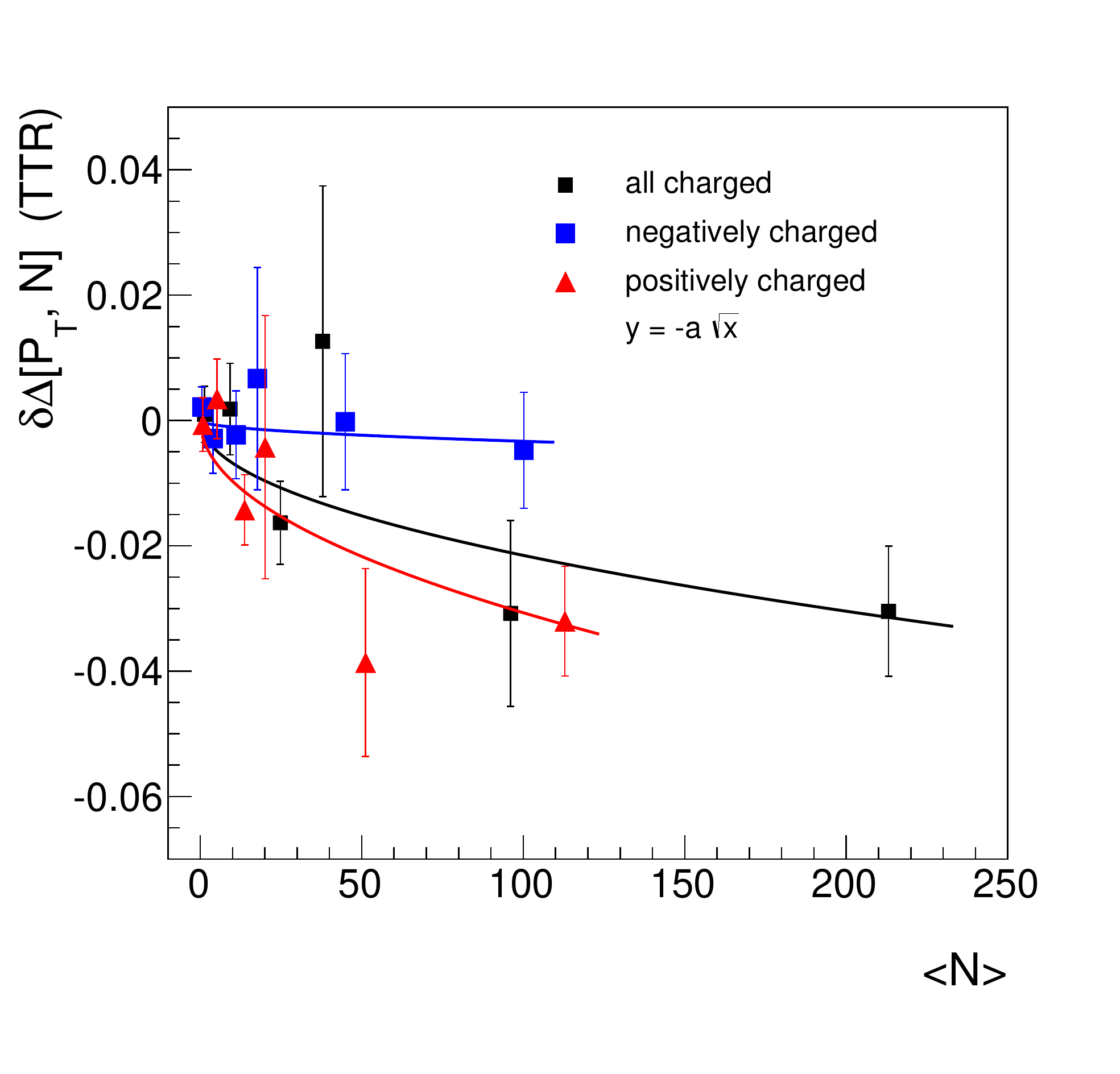}
\includegraphics[width=0.4\textwidth]{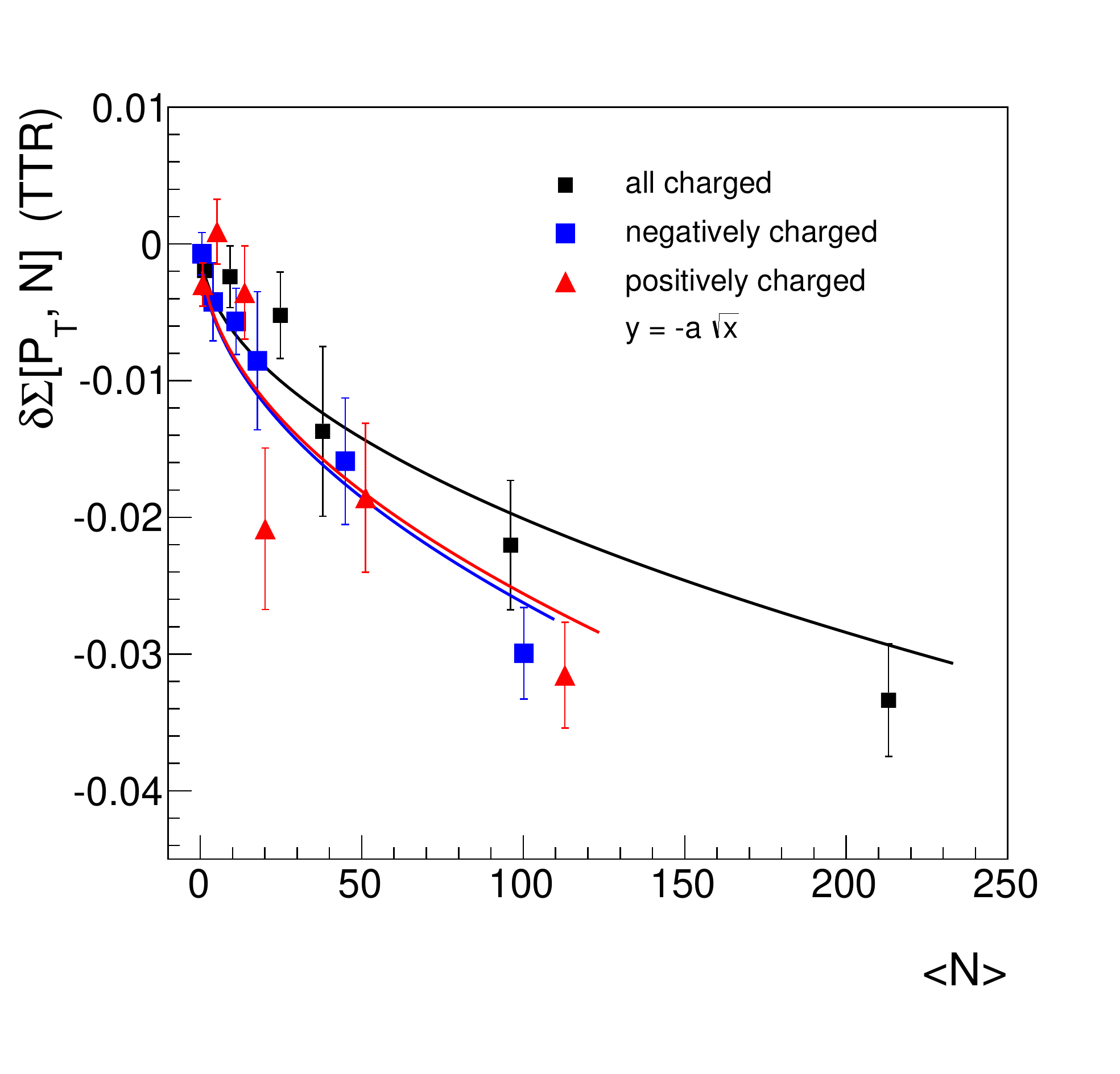}
\caption[]{(Color online) Additive corrections $\delta\Delta[P_T, N]$ (left) and $\delta\Sigma[P_T, N]$ (right) for limited two track resolution versus mean multiplicity $\langle N \rangle$ of accepted particles for $p+p$, C+C, Si+Si and three different centralities of Pb+Pb events at 158$A$ GeV/c. Estimates for positively charged, negatively charged and all charged particles are distinguished by different markers. The lines represent the analytical parametrization: 
$\delta\Delta (\Sigma)[P_T, N] (\langle N \rangle)=-a\sqrt{\langle N \rangle}$ with $a$ being the parameter of a fit to the data points.}
\label{delta_sigma_size_ttr}
\end{figure}

An additive correction for the limited two track resolution of the detector was applied to the values of $\Delta[P_T, N]$ and $\Sigma[P_T, N]$. The procedure to determine this correction was analogous to the one used to estimate the corrections for $\Phi_{p_T}$ in Refs.~\cite{Anticic:2003fd, Anticic:2008aa}. Mixed events were prepared for each of the analyzed data sets and then processed by the NA49 simulation software. The resulting simulated raw data were reconstructed and the measures $\Delta[P_T, N]$ and $\Sigma[P_T, N]$ were calculated using the same selection cuts as used for the real events. The additive two track resolution corrections 
$\delta\Delta[P_T, N]$ and $\delta\Sigma[P_T, N]$) were calculated as the difference between the values of $\Delta[P_T, N]$ (or $\Sigma[P_T, N]$) after detector simulation and reconstruction and before this procedure. The resulting corrections for the data of the energy scan are plotted in Fig.~\ref{delta_sigma_energy_ttr} and those for the data of the system size study in Fig.~\ref{delta_sigma_size_ttr}.

The statistical uncertainties on $\Delta[P_T, N]$ and $\Sigma[P_T, N]$ were obtained via the sub-sample method \cite{Anticic:2003fd, Anticic:2008aa}. The systematic uncertainties were estimated by varying event and track cut parameters (the procedures were identical to those applied for $\Phi_{p_T}$ in Refs.~\cite{Anticic:2003fd, Anticic:2008aa}).

%%%%%%%%%%%%%%%%%%%%%%%%%%%%%%%%%%%%%%%%%%%%%%%%%%%%%%%%%%%%%%%%%%%%
\section{Results and discussion}
\label{results}
%%%%%%%%%%%%%%%%%%%%%%%%%%%%%%%%%%%%%%%%%%%%%%%%%%%%%%%%%%%%%%%%%%%%

The results shown in this section refer to {\it accepted} particles, i.e., particles that are accepted by the detector and pass all kinematic cuts and track selection criteria as discussed in Sec.~\ref{datasets}. The data cover a broad range in $p_{T}$ ($0.005 < p_{T} < 1.5 $ GeV/c). The rapidity was restricted to the interval $1.1 < y^{*}_{\pi} < 2.6$ (forward rapidity) where contamination from beam produced $\delta$-rays is small. The selected azimuthal angle region is large and represents essentially the whole detector acceptance for the study of the system size dependence at 158$A$ GeV/c (see lines in Fig.~\ref{acc_sys_energy} (right)). It is more limited for the analysis of the energy dependence since the same region was chosen at all energies (see lines in Fig.~\ref{acc_sys_energy} (left)). Results are {\it not} corrected for limited kinematic acceptance. Such a correction is not possible since it depends on the, in general, unknown correlation mechanism. Instead the limited acceptance should be taken into account in the model calculations when comparing to experimental results.  However, corrections for limited two track resolution of the NA49 detector were applied (see Sec.~\ref{datasets} and Refs.~\cite{Anticic:2003fd, Anticic:2008aa}). A possible bias due to particle reconstruction losses and contamination in the accepted kinematic region was estimated to be small and is included in the systematic uncertainty of the results.

\subsection{Energy scan for central Pb+Pb interactions}
%%%%%%%%%%%%%%%%%%%%%%%%%%%%%%%%%%%%%%%%%%%%%%%%%%%%%%%%

\begin{figure}
\centering
\includegraphics[width=0.6\textwidth]{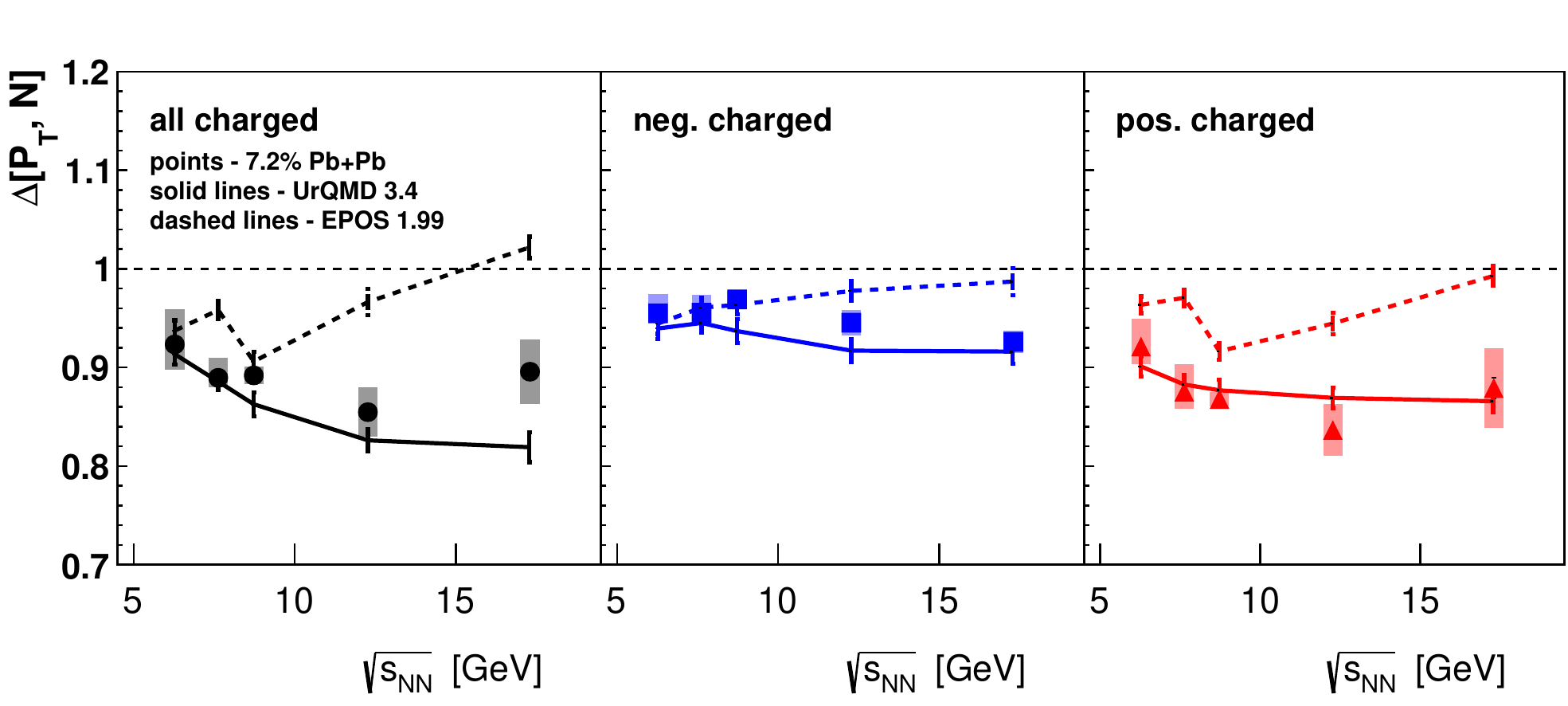}
\includegraphics[width=0.6\textwidth]{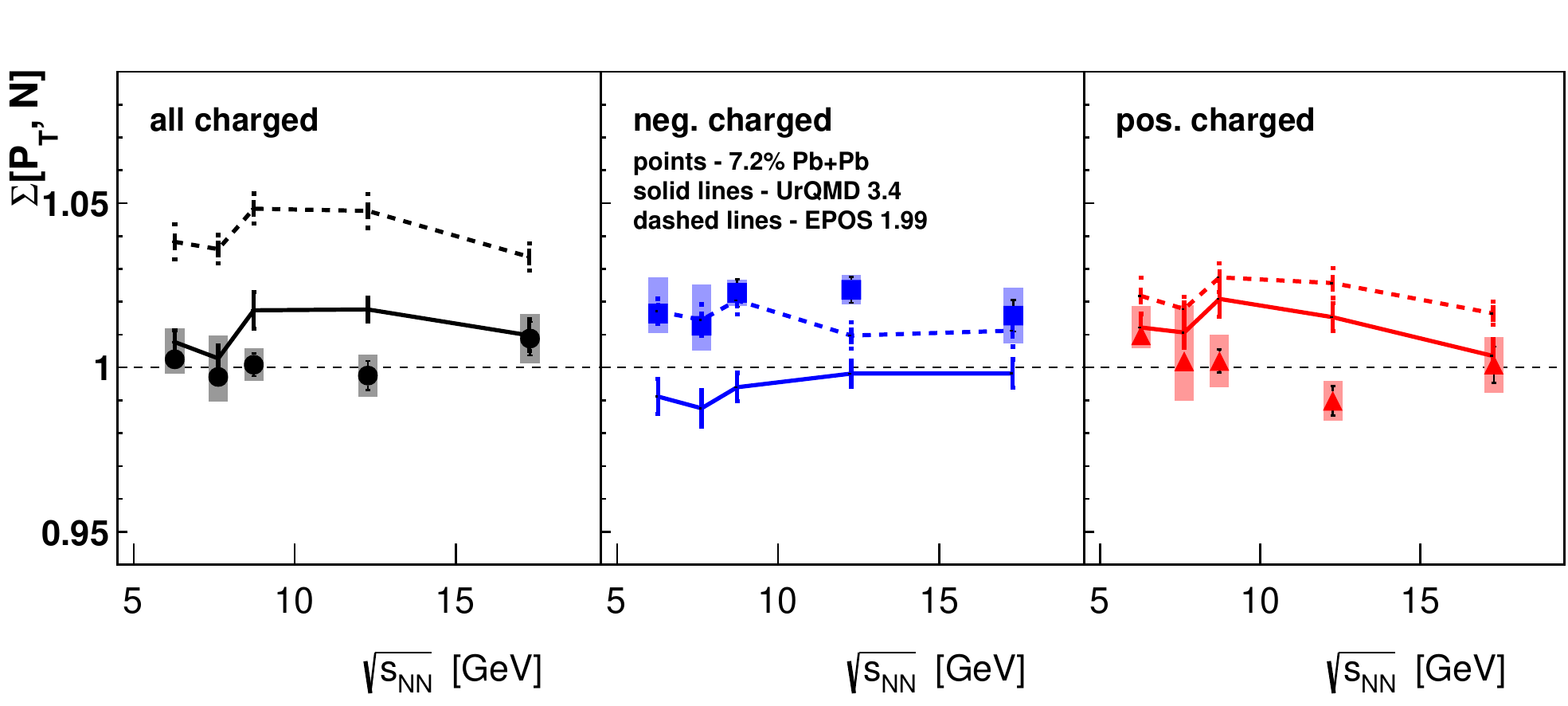}
\caption[]{(Color online) Energy dependence of $\Delta[P_T, N]$ (top) and $\Sigma[P_T, N]$ (bottom) for the 7.2\% most central Pb+Pb interactions. Statistical uncertainties are denoted by lines, systematic ones by color boxes. Data (points) are compared to predictions of the UrQMD 3.4 (solid lines) and EPOS 1.99 (dashed lines) models with acceptance restrictions as for the data. }
\label{final_results_energy_UrQMD}
\end{figure}

Figure \ref{final_results_energy_UrQMD} presents for the 7.2\% most central Pb+Pb interactions the energy dependence of the fluctuation measures $\Delta[P_T, N]$ and $\Sigma[P_T, N]$ calculated separately for all charged, negatively charged, and positively charged particles. The sample of negatively charged particles is composed mainly of $\pi^{-}$ mesons, whereas the sample of positively charged particles is dominated by $\pi^{+}$ mesons and protons. Therefore, the measured values of $\Delta[P_T, N]$ or $\Sigma[P_T, N]$ could differ between both charges. Moreover, among all charged particles additional sources of correlations could exist that are not present in positively or negatively charged particles separately. For all three charge selections the values of $\Delta[P_T, N]$ are smaller than one, the expectation for independent particle production. For $\Sigma[P_T, N]$ fluctuations for all and positively charged particles are close to the hypothesis of independent particle production (similar to the results on $\Phi_{p_{T}}$~\cite{Anticic:2008aa} which belongs to the same family of strongly intensive measures), whereas for negatively charged particles $\Sigma[P_T, N]$ values are higher than one. It was suggested in Refs.~\cite{Gorenstein:2013iua, Gazdzicki:2013ana} that values of $\Delta[P_T, N] < 1$ and $\Sigma[P_T, N] > 1$ can be explained as due to effects of Bose-Einstein statistics. Similarly, $\Phi_{p_{T}} > 0$ was predicted in Refs.~\cite{Mrowczynski:1999un, Mrowczynski:1998vt} as a consequence of Bose-Einstein correlations.

The measured values of $\Delta[P_T, N]$ and $\Sigma[P_T, N]$ are compared to predictions of the UrQMD~\cite{Bass:1998ca, Bleicher:1999xi} and EPOS~\cite{Werner:2008zza, Werner:2013yia} models in Fig.~\ref{final_results_energy_UrQMD} (solid and dashed lines respectively). The models do not simulate a phase transition or the critical point. However, resonance decays and effects of correlated particle production due to energy-momentum, charge and strangeness conservation laws are taken into account. The most central 7.2\% interactions were selected for comparison of the energy scan results, in accordance with the real NA49 events. The procedure of selecting the 7.2\% most central events was the following: a sample of minimum bias Pb+Pb events was produced. Then the distribution of the impact parameter $b$ was drawn and the value of $b$ was determined below which 7.2\% of the events remained. The resulting impact parameter range was $0 < b < 4.35$~fm in UrQMD and $0 < b < 4.00$~fm in EPOS. Finally, high statistics samples of UrQMD and EPOS events were produced in these impact parameter ranges separately for each energy. 

The measures $\Delta[P_T, N]$ and $\Sigma[P_T, N]$ were calculated from charged particles, consistent with originating from the main vertex. This means that mostly pions, protons, kaons and their anti-particles from the primary interaction were used because particles coming from the decays of $K^0_S$, $\Lambda$, $\Sigma$, $\Xi$, $\Omega$, etc. are suppressed by the track selection cuts. Therefore, the analyses of UrQMD and EPOS events were also carried out by using primary charged pions, protons, and kaons and their anti-particles. The tracking time parameter in the UrQMD model was set to 100 fm/c and therefore the list of generated kaons, pions and (anti-)protons did not contain the products of weak decays. In the parameter settings of the EPOS model the decays of $K^0_{S/L}$, $\Lambda$, $\Sigma$, $\Xi$, $\Omega$, etc. particles were explicitly forbidden. Finally, in the analysis of the UrQMD and EPOS events the same kinematic restrictions were applied as for the NA49 data.

Figure~\ref{final_results_energy_UrQMD}~(top) shows that the energy dependence of $\Delta[P_T, N]$ in the UrQMD model exhibits behavior similar to that observed in the measurements. In both cases one finds $\Delta[P_T, N] < 1$, i.e. values below those for independent particle production. As Bose-Einstein correlations are not implemented in the UrQMD model we conclude that in this model there must be another source(s) of correlation(s) leading to $\Delta[P_T, N] < 1$. The EPOS model shows $\Delta[P_T, N]$ values which are significantly higher that those obtained from the NA49 data and UrQMD. The comparisons for $\Sigma[P_T, N]$ can be seen in Fig.~\ref{final_results_energy_UrQMD}~(bottom). Here the predictions of UrQMD lie above the measurements for all charged and positively charged particles, whereas they are significantly below the results for negatively charged particles. On the other hand EPOS calculations for negatively charged particles are close to the data, but exceed the measurements even more than the UrQMD predictions for  all charged and positively charged particles.

The measured energy dependences of $\Delta[P_T, N]$ and $\Sigma[P_T, N]$ do not show any anomalies which might be attributed to  approaching the phase boundary or the critical point. However, it should be noted that due to the limited acceptance of NA49 and the additional restrictions used for this analysis the sensitivity for such fluctuations may be small if the underlying range of correlations in momentum space is large.

\subsection{System size dependence at 158$A$ GeV/c}
%%%%%%%%%%%%%%%%%%%%%%%%%%%%%%%%%%%%%%%%%%%%%%%%%%%%%%%%%%%%%%%%%%%%%%%%

\begin{figure}
\centering
\includegraphics[width=0.6\textwidth]{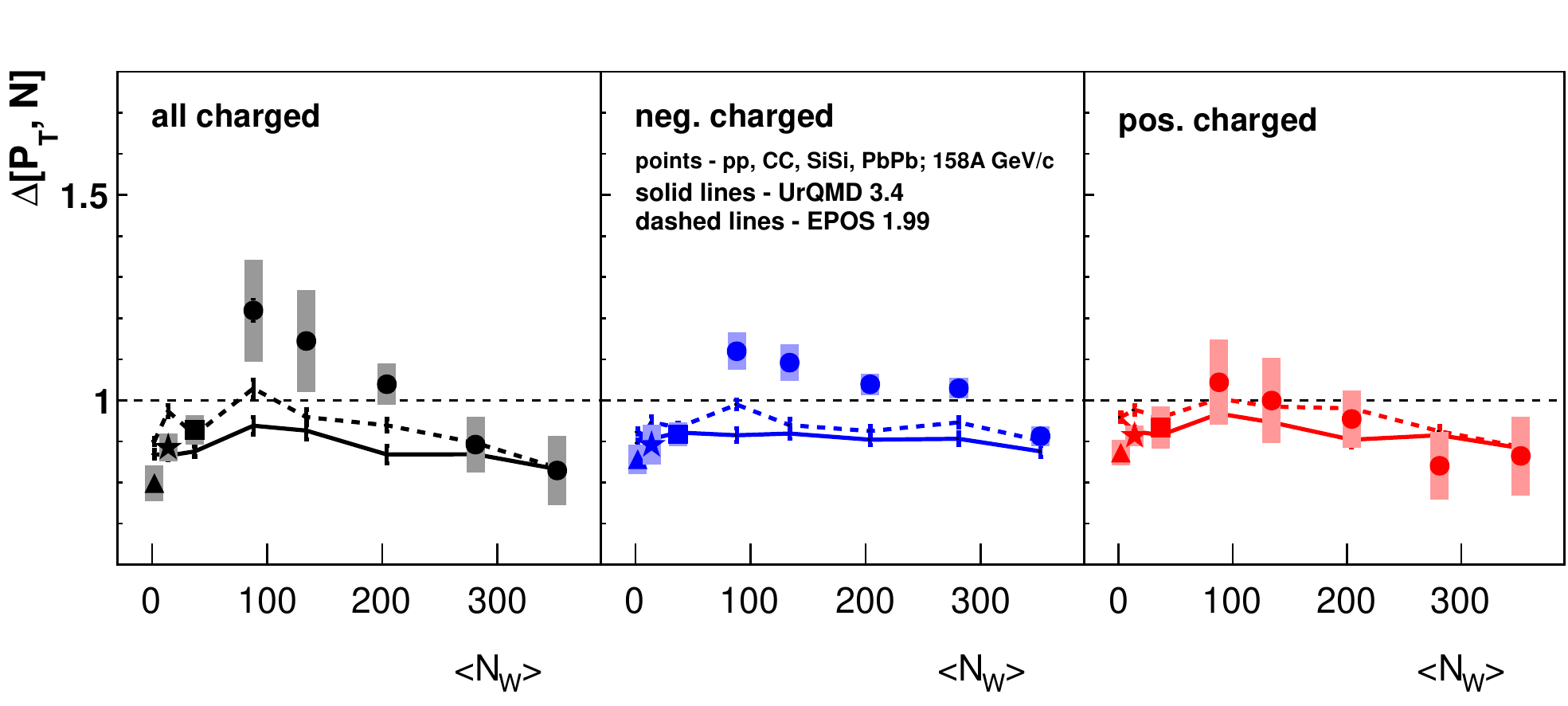}
\includegraphics[width=0.6\textwidth]{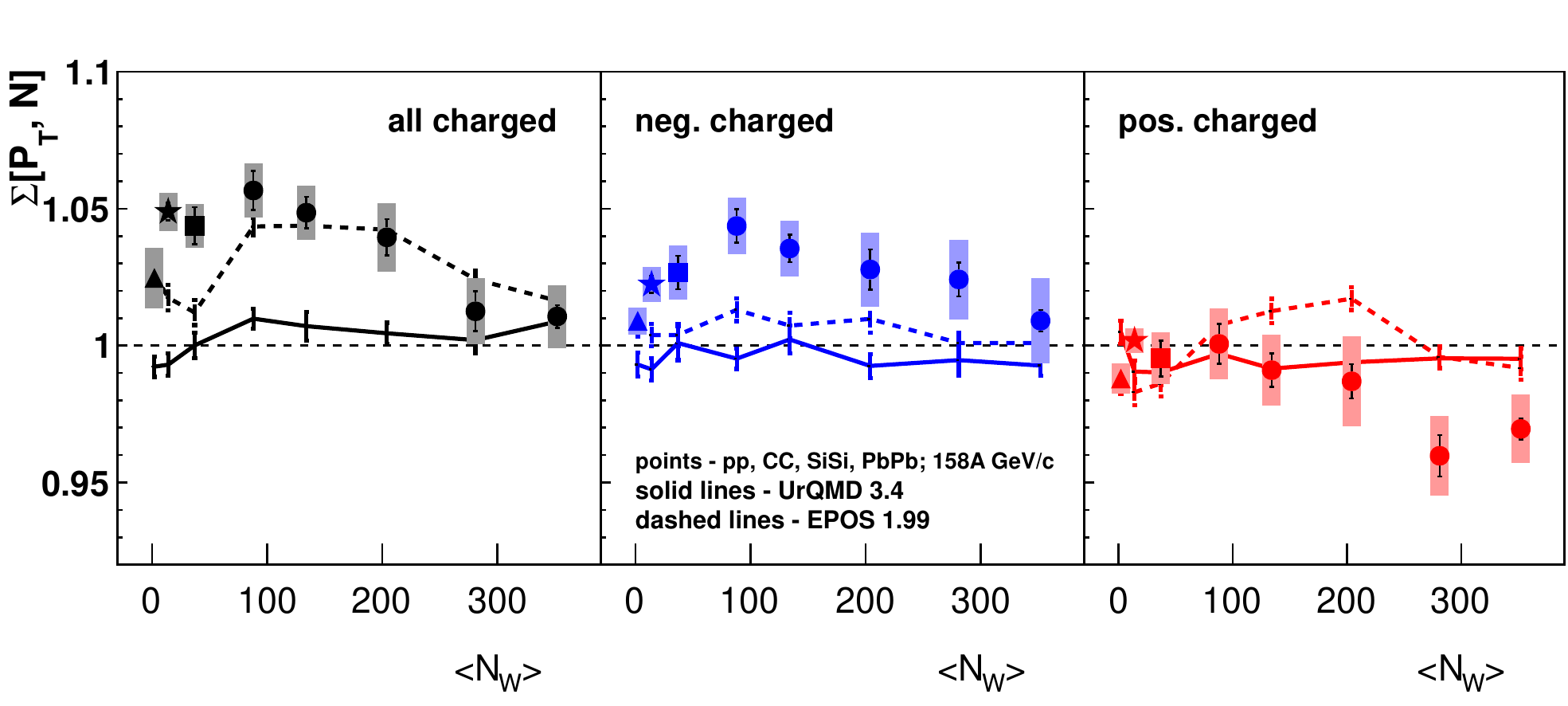}
\caption[]{(Color online) Dependence of $\Delta[P_T, N]$ (top) and $\Sigma[P_T, N]$ (bottom) on the mean number of wounded nucleons $\langle N_W \rangle$ on the size of the colliding nuclei ($p$ - triangles, C - stars, Si - squares, Pb - circles) and the centrality of Pb+Pb interactions at 158$A$ GeV/c. Statistical uncertainties are denoted by error bars, systematic uncertainties by colored boxes. Data (points) are compared to predictions of the UrQMD 3.4 (solid lines) and EPOS 1.99 (dashed lines) models with acceptance restrictions as for the data.}
\label{final_results_size_UrQMD}
\end{figure}

Figure \ref{final_results_size_UrQMD} presents the dependence of $\Delta[P_T, N]$ and $\Sigma[P_T, N]$ at 158$A$ GeV/c on the size of the colliding nuclei as well as on the centrality of Pb+Pb interactions. The measured values for all accepted charged particles and also for positively and negatively charged particles separately are plotted versus the mean number of wounded nucleons. The values of $\Delta[P_T, N]$ for $p+p$, C+C, Si+Si, and the two most central classes of Pb+Pb collisions are lower than one. For the more peripheral Pb+Pb interactions $\Delta[P_T, N]$ increases above one to a maximum for the most peripheral Pb+Pb collisions. The values of $\Sigma[P_T, N]$ for negatively and all charged particles are significantly above unity (the prediction of the independent particle production model) and also reach a maximum in the most peripheral Pb+Pb interactions. For positively charged particles the values are close to zero or below. The same behavior was observed for the measure $\Phi_{p_{T}}$~\cite{Anticic:2003fd} ($\Phi_{p_{T}}$ and $\Sigma[P_T, N]$ belong to the same family of strongly intensive measures). Finally, it is worth recalling that also for multiplicity fluctuations a maximum was observed in peripheral Pb+Pb collisions by NA49~\cite{Alt:2006jr}.

Figure \ref{final_results_size_UrQMD} shows that $\Delta[P_T, N]$ and $\Sigma[P_T, N]$ for all charged particles are usually higher than for either negatively or positively charged particles. Moreover, in case of $\Sigma[P_T, N]$, the values for positively charged particles are always lower than those for the negatively charged particles. In Ref.~\cite{Anticic:2003fd} it was shown that also values of $\Phi_{p_{T}}$ for positively charged particles were lower than those for negatively charged and for all charged particles. However, the same effect was observed in simulations using the HIJING model and the fact that $\Phi_{p_{T}}$ values for positively charged particles were always lower than those for negatively charged ones was found to be related to the limited acceptance and treatment of protons as pions in the calculation of rapidity.  

To further investigate the nature of the correlations leading to the observed values of the measures $\Delta[P_T, N]$ and $\Sigma[P_T, N]$  a toy model was constructed, in which an anti-correlation between mean transverse momentum per event ($P_T/N$) versus multiplicity ($N$) was assumed~\cite{Gorenstein:2013nea}. The parametrization of $P_T/N$ versus $N$ was taken from the NA49 $p+p$ data~\cite{Anticic:2003fd} (in the current paper the same $p+p$ data are used), resulting in $\Delta[P_T, N] = 0.816(0.005)$ and $\Sigma[P_T, N] = 1.008(0.002)$. This shows that Bose-Einstein correlations are not the only candidate for the explanation of $\Delta[P_T, N] < 1$ and $\Sigma[P_T, N] > 1$~\cite{Gorenstein:2013iua, Gazdzicki:2013ana}, but this observation, especially for smaller systems, may also be explained as due to the known $P_T/N$ versus $N$ anti-correlation~\cite{Gorenstein:2013nea}.

The system size dependences of $\Delta[P_T, N]$ and $\Sigma[P_T, N]$ were also compared to predictions of the UrQMD and EPOS models (the procedure of selecting the proper impact parameter range was analogous to that used in the case of the energy scan). Figure~\ref{final_results_size_UrQMD} shows that in the UrQMD model the values of $\Delta[P_T, N]$ and $\Sigma[P_T, N]$ are only weakly dependent on $\langle N_W \rangle$. For $\Delta[P_T, N]$ the values are slightly below independent particle production and for $\Sigma[P_T, N]$ they are close to 1. The pronounced maximum of fluctuations seen in the data at $\langle N_W \rangle \approx 90$ in Pb+Pb collisions is not reproduced by the UrQMD model. The EPOS model predictions for $\Delta[P_T, N]$ are similar to the predictions of UrQMD. Also the trends in $\Sigma[P_T, N]$ for negatively charged and positively charged particles are quite similar in UrQMD and EPOS. In contrast, for all charged particles the values of $\Sigma[P_T, N]$ are much higher in EPOS than in UrQMD and describe the NA49 results surprisingly well.

\subsection{Search for the critical point}

When searching for possible indications of a critical point it is most appropriate to plot the strength of fluctuations using the standard phase diagram coordinates temperature $T$ and baryochemical potential $\mu_B$. Moreover, central collisions of nuclei provide the cleanest interaction geometry. For such reactions fits of the hadron gas model (see e.g. Ref.~\cite{Becattini:2005xt}) were performed to determine the temperature $T_{chem}$ and baryochemical potential $\mu_B$ of the produced particle composition. These values are believed to be close to those of the hadronization along the transition line in the phase diagram. The value of $T_{chem}$ was found to decrease somewhat for collisions of larger nuclei, whereas $\mu_B$ decreases rapidly with collision energy.  

\begin{figure}
\centering
\includegraphics[width=0.6\textwidth]{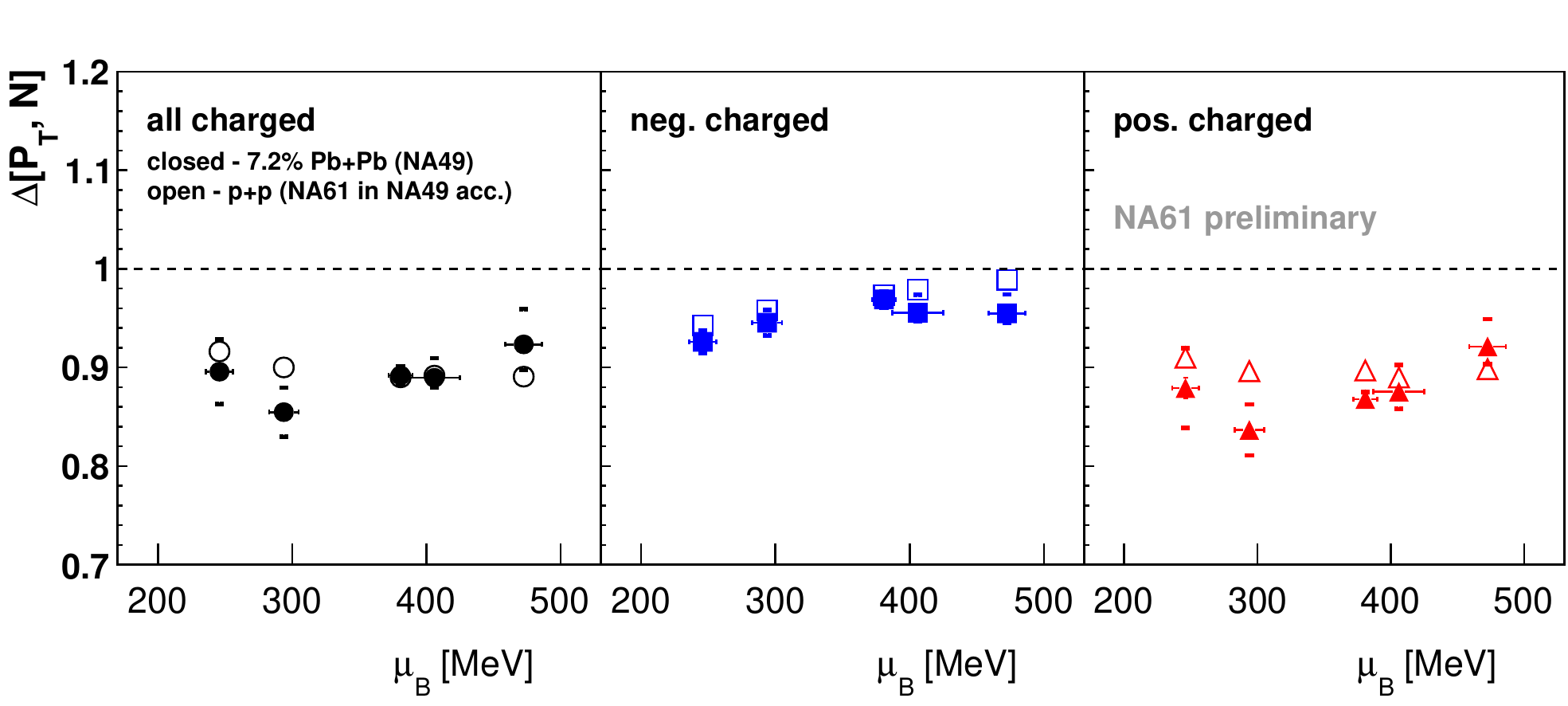}
\includegraphics[width=0.6\textwidth]{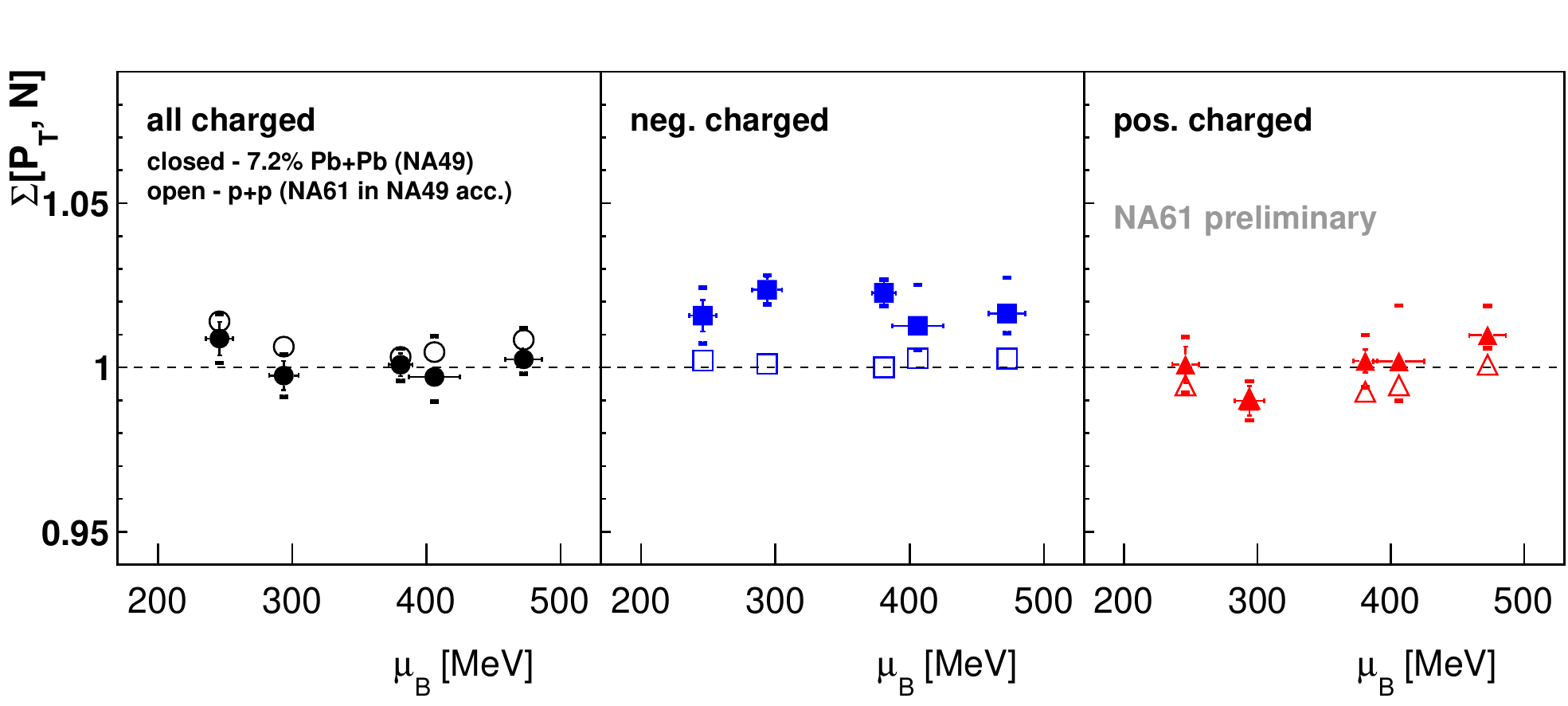}
\caption[]{(Color online) Energy ($\mu_B$) dependence of $\Delta[P_T, N]$ (top) and $\Sigma[P_T, N]$ (bottom) for the 7.2\% most central Pb+Pb interactions and a comparison to NA61 inelastic $p+p$ interactions. $\mu_B$ values for Pb+Pb collisions are taken from Ref.~\cite{Becattini:2005xt}. NA49 data indicate \cite{Becattini:2005xt} that at the top SPS energy $\mu_B$ does not depend on the system size (C+C, Si+Si, Pb+Pb). Therefore, the $\mu_B$ values for $p+p$ are also displayed and assumed to be the same as for Pb+Pb. NA61 data were taken from Refs.~\cite{Seyboth:2014ega, KG_Kielce_2013}. For NA61 only statistical uncertainties are shown.}
\label{miub_na49_na61}
\end{figure}

Results for $\Delta[P_T, N]$ and $\Sigma[P_T, N]$ for inelastic $p+p$ as well as central Pb+Pb collisions are shown in Fig.~\ref{miub_na49_na61} versus $\mu_B$. The $p+p$ results from NA61~\cite{Seyboth:2014ega, KG_Kielce_2013}, plotted for comparison, were obtained using the NA49 acceptance cuts. One observes little dependence on $\mu_B$ for both Pb+Pb or $p+p$ collisions. In particular, there is no indication of a maximum that might be attributed to the critical point. A similar conclusion was reached from the $\mu_B$ dependence of $\Phi_{p_{T}}$~\cite{Anticic:2008aa}. The measurements of $\Delta[P_T, N]$ are consistent for the two reactions. The values of $\Sigma[P_T, N]$ are close to unity with the exception of the higher result in Pb+Pb for negatively charged particles.   

\begin{figure}
\centering
\includegraphics[width=0.6\textwidth]{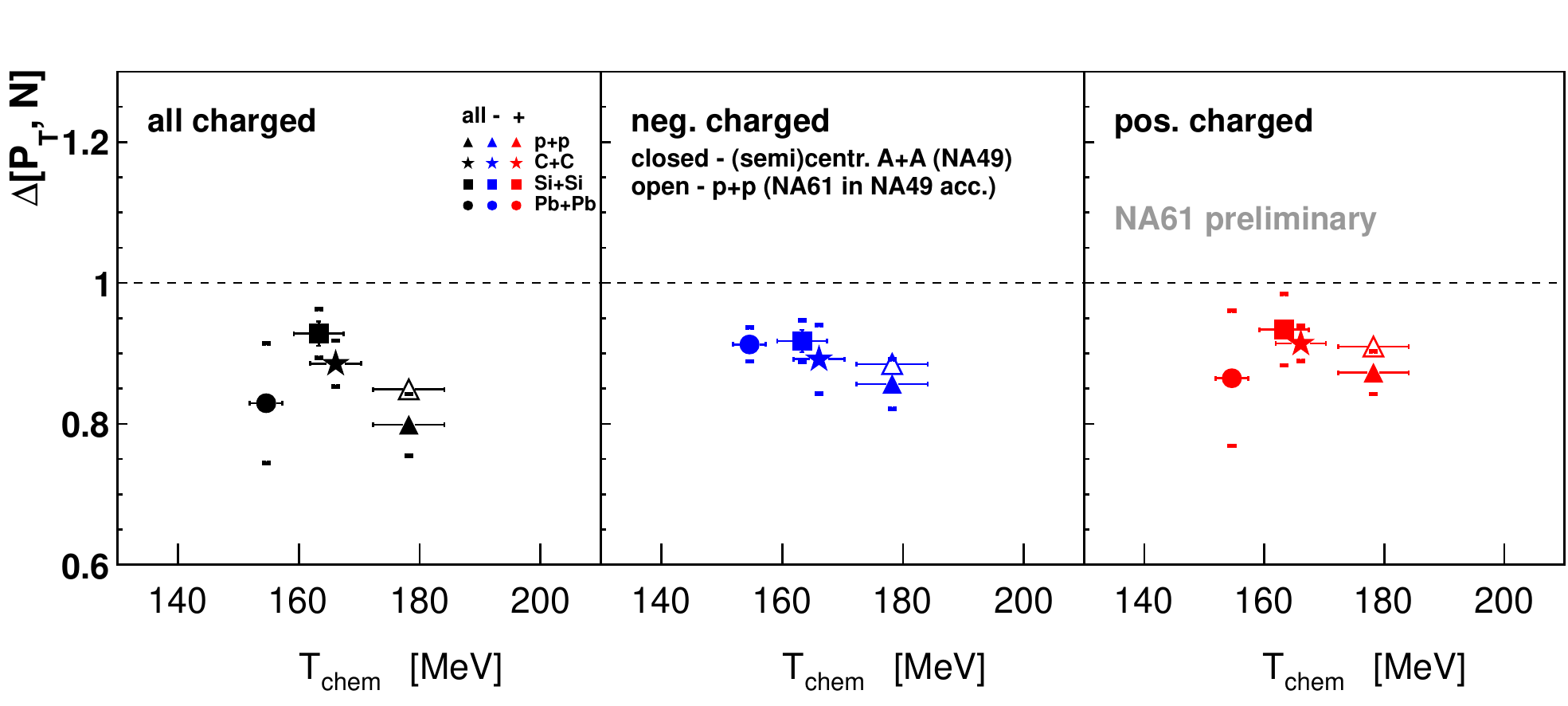}
\includegraphics[width=0.6\textwidth]{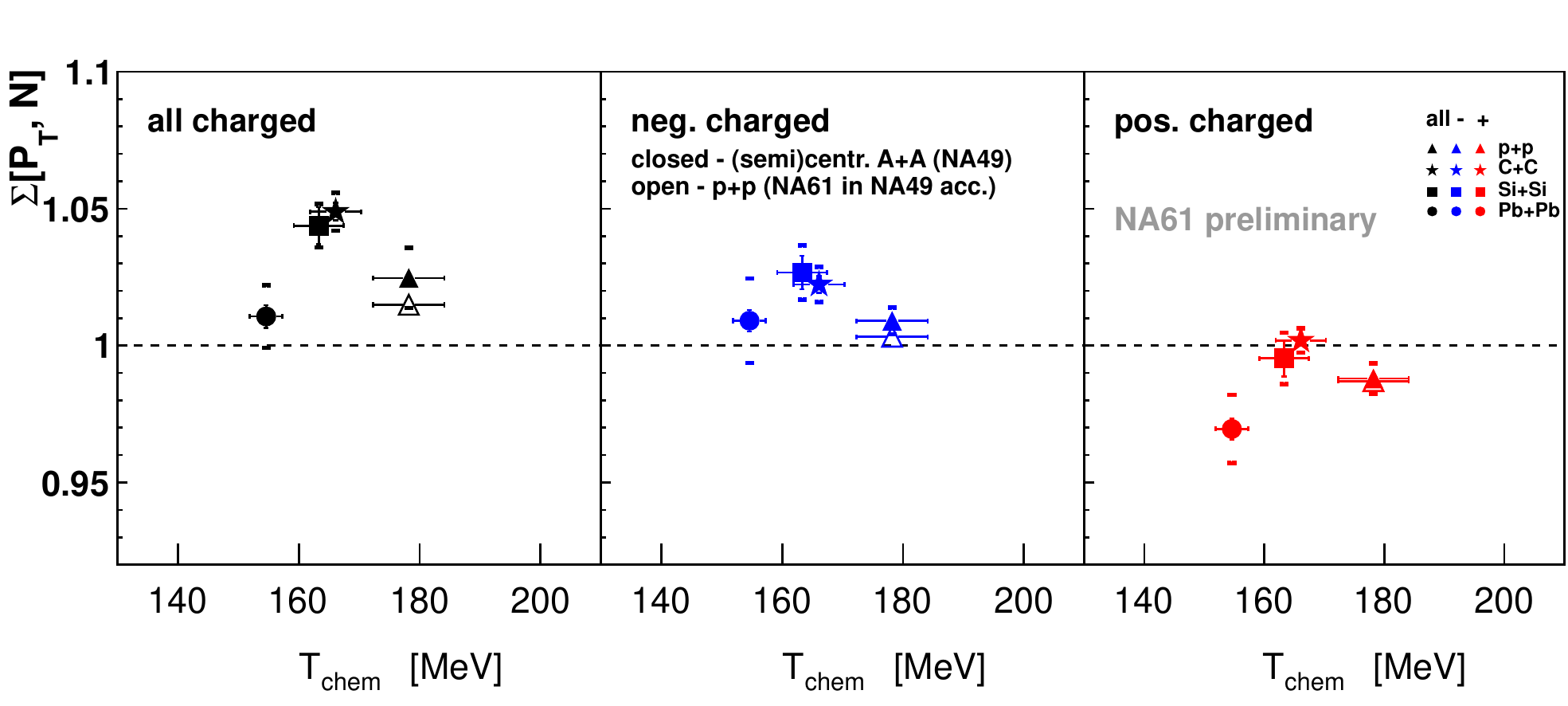}
\caption[]{(Color online) System size ($T_{chem}$) dependence of $\Delta[P_T, N]$ (top) and $\Sigma[P_T, N]$ (bottom) for (semi)central $A+A$ collisions at 158$A$ GeV/c and a comparison to NA61 inelastic $p+p$ interactions. $T_{chem}$ values for $p+p$, C+C, Si+Si and most central Pb+Pb collisions at 158$A$ GeV/c are taken from Ref.~\cite{Becattini:2005xt}. NA61 data were taken from Refs.~\cite{Seyboth:2014ega, KG_Kielce_2013}. For NA61 only statistical uncertainties are shown.} 
\label{Tch_na49_na61}
\end{figure}

The dependence of $\Delta[P_T, N]$, and $\Sigma[P_T, N]$ on $T_{chem}$ is shown in Fig.~\ref{Tch_na49_na61} at the beam momentum of 158$A$ GeV/c for $p+p$, semi-central C+C, Si+Si and central Pb+Pb reactions. The results for $p+p$ from NA49 (solid triangles) and NA61 (open triangles) are consistent.
% although NA61 used different experimental procedures. 
A maximum is observed for Si+Si interactions similar to the one found previously for $\Phi_{p_T}$ in Ref.~\cite{Anticic:2003fd}. There it was interpreted as a possible effect of the critical point~\cite{Grebieszkow:2009jr} consistent with QCD-based predictions of Ref.~\cite{Stephanov:1999zu, MS_private}. Interestingly, for the same system, studies of intermittency in the production of low mass $\pi^+\pi^-$ pairs~\cite{Anticic:2009pe} and of protons~\cite{Anticic:2012xb} found indications of power-law behavior with exponents that were consistent with QCD predictions for a CP. 

Unfortunately, theoretical predictions, for fluctuations at CP, are not yet published for the new fluctuation measures $\Delta[P_T, N]$ and $\Sigma[P_T, N]$. However, calculations for Si+Si collisions at 158$A$ GeV/c by using the Critical Monte Carlo (CMC) model~\cite{Antoniou:2000ms, Antoniou:2005am} are currently under study.

%%%%%%%%%%%%%%%%%%%%%%%%%%%%%%%%%%%%%%%%%%%%%%%%%%
\section{Summary}
%%%%%%%%%%%%%%%%%%%%%%%%%%%%%%%%%%%%%%%%%%%%%%%%%%

This paper reports on the continuing search at the CERN SPS by the NA49 experiment for evidence of the critical point of strongly interacting matter expected as a maximum of fluctuations. Results are presented on transverse momentum and multiplicity fluctuations of charged particles, produced at forward rapidities ($1.1 < y^{*}_{\pi} < 2.6$) in central Pb+Pb interactions at beam momenta 20$A$, 30$A$, 40$A$, 80$A$, and 158$A$ GeV/c, as well as in different systems ($p+p$, C+C, Si+Si, and Pb+Pb) at 158$A$ GeV/c. New strongly intensive measures of fluctuations, $\Delta[P_T, N]$ and $\Sigma[P_T, N]$, were measured. This paper is an extension of previous NA49 studies \cite{Anticic:2003fd, Anticic:2008aa} where the strongly intensive measure $\Phi_{p_T}$ was used to determine transverse momentum fluctuations. The quantities $\Delta[P_T, N]$ and $\Sigma[P_T, N]$ are dimensionless and have two reference values, namely they are equal to zero in case of no fluctuations ($P_T=const.$, $N=const.$) and one in case of independent particle production. Therefore, $\Delta[P_T, N]$ and $\Sigma[P_T, N]$ are preferable to $\Phi_{p_T}$ for which only one reference value is defined, i.e. $\Phi_{p_T}=0$ MeV/c for the model of independent particle production (IPM).

The NA49 results show no indications of a maximum in the energy dependence of transverse momentum (see also Ref.~\cite{Grebieszkow:2009jr}) and previously measured multiplicity \cite{Grebieszkow:2009jr} fluctuations in central Pb+Pb collisions throughout the CERN SPS energy range (but finer steps in the scan would be desirable). At all energies the values of $\Delta[P_T, N]$ are below one, i.e. below the expectation from the IPM. The values of $\Sigma[P_T, N]$ are close to one for all charged and positively charged particles and slightly higher in case of negatively charged particles. The effect for negatively charged particles can probably be explained as due to Bose-Einstein statistics. For positively charged particles the interpretation is less clear because other sources of correlations (for example resonance decays) can contribute to the correlation measures.

The system size dependence of $\Delta[P_T, N]$ and $\Sigma[P_T, N]$ (and the related measure $\Phi_{p_T}$ \cite{Anticic:2003fd, Grebieszkow:2009jr}) at 158$A$ GeV/c shows maxima for Si+Si and peripheral Pb+Pb interactions (for $\Delta[P_T, N]$ this maximum is mostly seen for peripheral Pb+Pb interactions). In central collisions values of $\Delta[P_T, N]$ are lower than one, whereas values of $\Sigma[P_T, N]$ for all charged and negatively charged particles are higher than one (weak anticorrelations). The maximum of $\Sigma[P_T, N]$ for all charged particles in C+C and Si+Si interactions is about 5\% higher than the base line defined by the IPM. Also previously studied multiplicity fluctuations for the most central $A+A$ collisions were found to show a maximum for Si+Si reactions at 158$A$ GeV/c~\cite{Grebieszkow:2009jr}. The excess of transverse momentum and multiplicity fluctuations is two times higher for all charged than for negatively charged particles as expected for the CP \cite{Stephanov:1999zu}.  

The NA49 collaboration also searched for evidence of the critical point in an intermittency analysis of low-mass $\pi^+\pi^-$ pair~\cite{Anticic:2009pe} and proton~\cite{Anticic:2012xb} production. Indications of power-law behavior consistent with that predicted for a CP were found in the same Si+Si interactions at 158$A$ GeV/c. The intriguing results strongly motivate the ongoing critical point search by the successor experiment NA61/SHINE~\cite{shine} which performs a systematic two-dimensional scan (SPS energies and system size ($p$, Be, Ar, Xe, Pb)) of the phase diagram of strongly interacting matter. A maximum of several CP signatures, the so-called {\it hill of fluctuations}, would signal the existence of the CP. The RHIC Beam Energy Scan~\cite{Aggarwal:2010cw} pursues a complementary program measuring higher order moments and cumulants of net-charge and net-proton distributions in Au+Au collisions. So far no clear evidence for the CP was found \cite{Adamczyk:2014fia, Adamczyk:2013dal}. 
Thus the possible existence of the CP remains an interesting and challenging question.

%%%%%%%%%%%%%%%%%%%%%%%%%%%%%%%%%
%%%%%%%%%%%%%%%%%%%%%%%%%%%%%%%%%

\vspace{1.0cm}
{\bf Acknowledgments:}~This work was supported bythe US Department of Energy Grant DE-FG03-97ER41020/A000,
the Bundesministerium fur Bildung und Forschung (06F 137), Germany,
the German Research Foundation (grant GA 1480/2-2),
the National Science Centre, Poland (grants DEC-2011/03/B/ST2/02617, DEC-2011/03/B/ST2/02634, and DEC-2014/14/E/ST2/00018),
the Hungarian Scientific Research Foundation (Grants OTKA 68506, 71989, A08-77719 and A08-77815),
the Bolyai Research Grant,
the Bulgarian National Science Fund (Ph-09/05),
the Croatian Ministry of Science, Education and Sport (Project 098-0982887-2878)
and 
Stichting FOM, the Netherlands.

%%%%%%%%%%%%%%%%%%%%%%%%%%%%%%%%%%%%%%%%%%%%%%
%%%%%%%%%%%%%%%%%%%%%%%%%%%%%%%%%%%%%%%%%%%%%%

\end{document}